\documentclass[11pt]{article}
\usepackage{amsfonts,amssymb,eucal}
\usepackage{amsthm}
 \usepackage{amsmath}
\usepackage{amscd}
 \usepackage{latexsym} 

\newcommand{\R}{\ensuremath{\mathbb{R}}}
\newcommand{\C}{\ensuremath{\mathbb{C}}}

\newcommand{\T}{\ensuremath{\mathbb{T}}}

\begin{document}

\title{\bf  Complex Hadamard matrices from Sylvester inverse orthogonal matrices}
\author{Petre Di\c t\u a\\
National Institute of Physics and Nuclear Engineering,\\
P.O. Box MG6, Bucharest, Romania\\ email: dita@zeus.theory.nipne.ro}

\maketitle

\begin{abstract}
A novel method to obtain  parametrisation of complex inverse orthogonal matrices is provided. These matrices are  natural generalizations of complex Hadamard matrices  which depend on non zero complex parameters.  The method we use is via doubling the size of inverse complex conference matrices.  When the free  parameters  take values on the unit circle the inverse orthogonal matrices transform into complex Hadamard matrices, and in this way we find new parametrisation of Hadamard matrices for dimensions $ n=8,\,10,$ and $12$.

\end{abstract}

\section{Introduction}

In the seminal paper  \cite{JJS} Sylvester defined the notion of a  {\em self-reciprocal matrix} as a square array of elements of which each is proportional to its first minor. If the sum of the squares of the terms in each row or in each column is equal to unity the matrix becomes strictly orthogonal. When the strictness condition was withdrawn he defined  a more general class of matrices by a homographic relation between each element and its first minor. However in this paper   we shall make use only   of the particular class of  inverse orthogonal matrices $A=(a_{ij})$ that are  those matrices whose inverse  is given by 
$A^{-1}=(1/a_{ij})^t=(1/a_{ji})$, where $t$ means transpose, and their entries $0\ne a_{ij}\in \C$  satisfy the relation
\begin{eqnarray}
A A^{-1}= n I_n \label{inv}
\end{eqnarray}
In the above relation $I_n$ is the $n$-dimensional identity matrix. When the entries  $a_{ij}$ take values on the unit circle $A^{-1}$ coincides with the Hermitian conjugate $A^*$ of $A$, and in this case (\ref{inv}) is the definition of  complex Hadamard matrices. Complex Hadamard matrices have applications in quantum information theory, several branches of combinatorics, digital signal processing, image analysis, coding theory, cryptology, etc. 

Complex orthogonal matrices also  appeared in the description of topological invariants of knots and links, see e.g. Ref.\,\cite{VFRJ}. They were called two-weight spin models  being related to symmetric statistical Potts models. These matrices  have been  generalized to two-weight spin models, \cite{KMW}, also called generalized spin models, by removing the symmetry condition. After that these models have been extended to four-weight spin models, \cite{BB}. These last  matrices are also known under the name of  type II matrices,
see Ref.\,\cite{KN}.
Particular cases of type II matrices also appeared as  generalized Hadamard transform for processing  multiphase or   multilevel signals, see \cite{LRP} and \cite{H}, which includes the classical Fourier, Walsh-Hadamard and Reverse Jacket transforms. They are defined for $2n\times 2n$-dimensional matrices and, until now,  only for  a particular class given in terms of   $p^{th}$ roots of unity and/or one complex non-zero parameter \cite{LRP}. The aim of the paper is to provide an analytic method for the construction of inverse orthogonal matrices, and, as a byproduct,  to get from them nonequivalent complex Hadamard matrices. It is well known, see \cite{KN},  that a complete classification of inverse orthogonal matrices was given only for dimensions, $n\le 5$. For  $n\ge 6$ the problem is still  open. Even the ``simpler'' problem of classification of nonequivalent complex Hadamard matrices is not yet solved, and new families frequently appear, see in this respect the  papers \cite{MRS}, \cite{BN}, \cite{MS}, \cite{FS} and \cite{FS2}.    The  method we use in the following to construct complex Hadamard matrices is via doubling the size of inverse complex conference matrices. The first non trivial case is $n=4$, and  here we give a few examples for dimensions $m= 2n$ when $n= 4,\,5,\,{\rm and}\,\,6$.

\section{Inverse Orthogonal Matrices}

 The complex  $n\times n$
 conference matrices, $ C_n$,  are matrices with $a_{ii}=0,\,\, i=1,\dots,n$ and
 $|a_{ij}|=1, \,i\ne j$ that  satisfy
\begin{eqnarray}
C_n\,C_n^* =(n-1)I_n\label{con}\end{eqnarray}
where $C_n^*$ is the Hermitian conjugate of $C_n$.
Conference matrices, $C_n$,  are important because by  construction the matrix
\begin{eqnarray}
H_{2n} = \left(\begin{array}{cc}
C_n +I_n&C_n^* -I_n\\
C_n -I_n&-C_n^* -I_n\end{array}\right)\label{conf}
\end{eqnarray}
  is  complex Hadamard of size $ 2n\times 2n$, as one can easily verify.
The above formula can be slightly generalized by taking the entries of the first column in (\ref{conf}) as $C_n \pm I_n e^{i a} $ and those from the second column as
$\pm C_n^* -I_n  e^{-i a}$, with $a \,\in$ \R.

Let us now suppose that we have a complex conference matrix $C_n$ which depends  on a few arbitrary phases $e^{i \alpha_j},\,\,j=1,...,k$, then it  can be transformed into a complex inverse orthogonal conference  matrix by the change $e^{i \alpha_j}\rightarrow a_j $ with $0\ne a_j\in$\,{\C} complex non-zero numbers. Of course one can look directly for  inverse orthogonal conference matrices, but our method provided us quite easy formulas for the cases $n=4,\,5,\,6$. Thus the  complex inverse  orthogonal conference matrices are  those matrices   with $a_{ij},\,i\ne j$, complex non-zero numbers, and $a_{ii}=0$, defined  by a similar relation to the relation (\ref{con}). It is well known that for complex Hadamard and conference matrices the Hermitian conjugate coincides with the inverse matrix. Hence  to extend the  above construction (\ref{conf}) to complex inverse orthogonal conference matrices we have to provide a recipe for a proper treating of  zero entries on the main diagonal.
 Our formula for   inverse is given by
\begin{eqnarray}
C_n^{-1}=(1/(C_n + I_n)-I_n)^t\label{W1}\end{eqnarray}
where  $1/A$ is the matrix whose elements are $1/a_{ij}$. We remark that the above relation makes sense since all the entries of $ C_n + I_n $ are complex non-null numbers.
In the case of complex orthogonal matrices the formula (\ref{conf}) takes the form
\begin{eqnarray}
O_{2n} = \left(\begin{array}{cc}
C_n +a\,I_n&C_n^{-1} -I_n/a\\
C_n -a\,I_n&-C_n^{-1} -I_n/a\end{array}\right)\label{orth}
\end{eqnarray}
with $a\in$\,\C$^*$  any complex  non-zero number. From the equations  (\ref{W1}) and (\ref{orth}) it results that the matrix $O_{2n}$  could depend  nonlinear on the parameter $a$ and all the parameters entering $C_n$.

Now we show how to proceed in this case to obtain complex Hadamard matrices from Sylvester orthogonal matrices.   First we consider the case $n=4$ and start with the complex  conference matrix given in \cite{D}, p. 5373, by changing its arbitrary phase $e^{i t}$ into the complex number $b\ne 0$
\begin{eqnarray}
C_4=\left( \begin{array}{cccc}
0&1&1&1\\
1&0&-b&b\\
1&b&0&-b\\
1&-b&b&0
\end{array}\right)\label{conf4}
\end{eqnarray}
 We remark 
 that in the doubling formula (\ref{orth}) the diagonal matrix $I_n$ acts as a symmetry breaking and by consequence we can multiply the columns, or the rows, of $C_n$  by arbitrary non-null complex numbers. Thus in the above case and those similar to  we multiply (\ref{conf4}) at right by a diagonal matrix. When $n=4$ the diagonal of this diagonal matrix has the form $d_1=(A_1,A_2,A_3,A_4)$, where $A_i$  are non-zero complex numbers, and by taking into account the nonlinear dependence of the formula (\ref{orth}) on $C_n$ entries  we will   obtain   complex orthogonal matrices  depending on more free parameters. In this case from (\ref{orth}) one gets
\begin{eqnarray}
 O_8=
\left[\begin{array}{cccccccc}
a&A_2&A_3&A_4&-\frac{1}{a}&\frac{1}{A1}&\frac{1}{A1} &\frac{1}{A1} \\*[2mm]
A_1&a&- A_3 b& A_4b&\frac{1}{A_2}&-\frac{1}{a}&\frac{1}{A_2b}&-\frac{1}{A_2b} \\*[2mm]
A_1& A_2b&a&- A_4b&\frac{1}{A_3}&-\frac{1}{A_3 b}&-\frac{1}{a}&\frac{1}{A_3 b}\\*[2mm]
A_1&-A_2b&A_3b&a&\frac{1}{A_4}&\frac{1}{A_4 b}&-\frac{1}{A_4b}&-\frac{1}{a}\\*[2mm]
-a&A_2&A_3&A_4&-\frac{1}{a}&-\frac{1}{A_1}&-\frac{1}{A_1} &-\frac{1}{A_1} \\*[2mm]
A_1&-a&-A_3 b&a\, A_4b&-\frac{1}{A_2}&-\frac{1}{a}&-\frac{1}{A_2b}&\frac{1}{A_2b}\\*[2mm]
A_1&A_2b&-a&-A_4b&-\frac{1}{A_3}&\frac{1}{A_3 b}&-\frac{1}{a}&-\frac{1}{A_3b}\\*[2mm]
A_1&-A_2b&A_3b&-a&-\frac{1}{A_4}&-\frac{1}{A_4 b}&\frac{1}{A_4 b}&-\frac{1}{a}
\end{array}\right]\label{d8}
\end{eqnarray}

 Now we multiply  $O_8$ matrix (\ref{d8}) at right and, respectively, at left by diagonal matrices generated by the inverse of the first row, respectively, of the first column, and make  the substitutions
\begin{eqnarray} A_1= 1/c,\,\,A_2= 1/d ,\,\,A_3= 1/e,\,\, A_4= 1/f\end{eqnarray}
 After interchanging  through a transposition, the fifth and eight rows one gets the following form of the $O_8$ matrix 
\begin{eqnarray}
 O_8=
\left[\begin{array}{cccccccc}
1&1&1&1&1&1&1&1\\
1&a^2 c\, d&-a b c&a b c &-a^2 c\, d&-1&\frac{a d}{b}&\frac{-a d}{b}\\*[2mm]
1&a b c &a^2 ce &-a b c&-a^2 c e&\frac{-a e}{b}&-1&\frac{a e}{b}\\*[2mm]
1&-a b c&a b c &a^2c f&-a^2c f&\frac{a f}{b}&\frac{-a f}{b}&-1\\*[2mm]
1&-abc&abc&-a^2cf&a^2cf&\frac{-af}{b}&\frac{af}{b}&-1\\*[2mm]
1&-a^2 c d&-abc&abc& a^2 c d&-1&\frac{-ad}{b}&\frac{ad}{b}\\*[2mm]
1&abc &-a^2c e&-abc&a^2 ce&\frac{ae}{b}&-1&\frac{-ae}{b}\\
1&-1&-1&-1&-1&1&1&1
\end{array}\right]\label{O8}
\end{eqnarray}
matrix which apparently  depends on six complex arbitrary non-zero parameters. This does not happen as F  Sz\"oll\H{o}si pointed out to us \cite{ FS1}. In fact the above matrix depends on {\it four} arbitrary parameters which are obtained by the following substitutions in (\ref{O8})
  \begin{eqnarray}abc \rightarrow a, \;ad/b \rightarrow b, \;ae/b \rightarrow c, \; {\rm and}\; af/b \rightarrow d \label{dd}\end{eqnarray}
and  the final  form is
\begin{eqnarray}
 O_8^{(4)}=
\left[\begin{array}{cccccccc}
1&1&1&1&1&1&1&1\\
1&ab&-a&a&-ab&-1&b&-b\\
1&a&ac&-a&-ac&-c&-1&c\\
1&-a&a&ad&-ad&d&-d&-1\\
1&-a&a&-ad&ad&-d&d&-1\\
1&-ab&-a&a&ab&-1&-b&b\\
1&a&-ac&-a&ac&c&-1&-c\\
1&-1&-1&-1&-1&1&1&1
\end{array}\right]\label{O8a}
\end{eqnarray}
By the above re-parametrization (\ref{dd}) the quadratic dependence on $a$ in (\ref{O8}) disappears in the final form  (\ref{O8a}).

 When the new parameters defined by (\ref{dd}) entering (\ref{O8a}) 
  take values on the  4-dimensional torus \T$^4$, i.e. when one make the substitutions $a\rightarrow e^{ ia}$, etc, $O_8$ gets a complex Hadamard matrix. It can be written in the form
\begin{eqnarray}
D_{8}^{(4)}(a,b,c,d)=
 H_{8}\circ  EXP\left( i\cdot R_8^{(4)}(a,b,c,d)\right)\label{affine1}
\end{eqnarray}
The above relation is the standard form introduced in \cite{TZ} to present complex Hadamard matrices, where now all the four parameters, $a,\,b,\,c,\,d \in \R$, are real numbers, $\circ$ denotes the Hadamard product, $ i=\sqrt{-1}$ and $H_{8}$ is obtained from (\ref{O8a}) by taking all the four parameters equal to unity, i.e.
\begin{eqnarray}
 H_{8}=\left[\begin{array}{rrrrrrrr}
1&1&1&1&1&1&1&1\\
1&1&-1&1&-1&-1&1&-1\\
1&1&1&-1&-1&-1&-1&1\\
1&-1&1&1&-1&1&-1&-1\\
1&-1&1&-1&1&-1&1&-1\\
1&-1&-1&1&1&-1&-1&1\\
1&1&-1&-1&1&1&-1&-1\\
1&-1&-1&-1&-1&1&1&1\end{array}\right]\label{H8}
\end{eqnarray}
and
\begin{eqnarray}
R_8^{(4)}=
\left[\begin{array}{cccccccc}
\bullet&\bullet&\bullet&\bullet&\bullet&\bullet&\bullet&\bullet\\
\bullet&a+b&a&a&a+b&\bullet&b&b\\
\bullet&a&a+c&a&a+c&c&\bullet&c\\
\bullet&a&a&a+d&a+d&d&d&\bullet\\
\bullet&a&a&a+d&a+d&d&d&\bullet\\
\bullet&a+b&a&a&a+b&\bullet&b&b\\
\bullet&a&a+c&a&a+c&c&\bullet&c\\
\bullet&\bullet&\bullet&\bullet&\bullet&\bullet&\bullet&\bullet
\end{array}\right]\label{R8}
\end{eqnarray}
where $\bullet$ means zero. It is well known that matrices under Hadamard product and usual addition generate a commutative algebra such that all the usual functions, like exponential entering  (\ref{affine1}), are well defined.

 The only previous known result that  depends on four free parameters is the matrix $S_8^{(4)}$, see \cite{MRS}. If one multiplies at left $O_8^{(4)}$ by the diagonal matrix $diag(1,\,1/ab\,,1/a\,,-1/a\,,-1/a,\,-1/ab\,,-1)$ followed by re-parametrization 
$(a\rightarrow 1/d,\,b\rightarrow -i/c,\, c\rightarrow -ia/d,\,d \rightarrow ib/d )$ one gets a matrix equivalent to  $S_8^{(4)}$ transpose, \cite{FS1}.  This shows that the equivalence problem of two matrices is not  easy to solve.

If in  relation (\ref{O8a}) one makes the replacements $a =  i,\,b=c=d=1 $ one gets the matrix 
\begin{eqnarray}
D_8=\left[\begin{array}{rrrrrrrr}
1&1&1&1&1&1&1&1\\
1& {\bf i}& {\bf -i}& {\bf i}& {\bf -i}&-1&1&-1\\
1&{\bf i}& {\bf i}& {\bf -i}& {\bf -i}&-1&-1&1\\
1&{\bf -i}& {\bf i}& {\bf i}& {\bf -i}&1&-1&-1\\
1&{\bf -i}& {\bf i}& {\bf -i}& {\bf i}&-1&1&-1\\
1&{\bf -i}& {\bf -i}& {\bf i}& {\bf i}&-1&-1&1\\
1&{\bf i}& {\bf -i}& {\bf -i}& {\bf i}&1&-1&-1\\
1&-1&-1&-1&-1&1&1&1\end{array}\right]\label{g8}
\end{eqnarray}

Similar matrices are the $K_4$ matrix obtained by Horadam, \cite{H}, from a back-circulant matrix at its turn derived from a quadriphase perfect sequence, the $S_8$ matrix obtained  by Matolcsi {\em et al}, \cite{MRS},  via tiling abelian groups, and the $J_8$ matrix obtained by Lee and Vavrek, \cite{LV},  from a modified Paley construction. All these four matrices are nonequivalent.  $D_8$  has the form of a jacket matrix, i.e. its  entries from the first row and column are 1, while  the entries from the last row and column are  $\pm 1$. 

If now in (\ref{O8a}) one  takes  $a=1$ and $b=c=d=i $  one gets an other matrix similar to (\ref{g8}) that is not a jacket matrix 
\begin{eqnarray}
d_8=\left[\begin{array}{rrrrrrrr}
1&1&1&1&1&1&1&1\\
1&{\bf i}&-1&1&-{\bf i}&-1&{\bf i}&-{\bf i}\\
1&1&{\bf i}&-1&-{\bf i}&-{\bf i}&-1&{\bf i}\\
1&-1&1&{\bf i}&-{\bf i}&{\bf i}&-{\bf i}&-1\\
1&-1&1&- {\bf i}&{\bf i}&-{\bf i}&{\bf i}&-1\\
1&-{\bf i}&-1&1&{\bf i}&-1&-{\bf i}&{\bf i}\\
1&1&-{\bf i}&-1&{\bf i}&{\bf i}&-1&-{\bf i}\\
1&-1&-1&-1&-1&1&1&1
\end{array}\right]\label{j8}
\end{eqnarray}

By using the doubling formula
\begin{eqnarray}
C=\left( \begin{array}{cc}
A&D B\\
A&-D B\end{array}\right)\label{h16}
\end{eqnarray}
 see \cite{D}, Eq. (8), where $A$ and $ B$ are matrices of the form $D_{8}^{(4)}$ each one depending on four different parameters, and $D\in$ \T$^8$ is a diagonal matrix with its first entry 1, and  the other entries  seven arbitrary phases, one gets a matrix  $D_{16}^{(15)}$ that depends on 15 arbitrary parameters.
The  known results for this dimension are the matrices $F_{16}^{(17)}$ from 
\cite{TZ}, and $R_{16}^{(11)}$ from \cite{MRS}, which depend on 17, respectively 11, free parameters. 

When $n=5$ we start  with the complex orthogonal conference matrix
\begin{eqnarray}
C_5=\left[\begin{array}{ccccc}
0&1&1&1&1\\
1&0&b&\omega\, b&\omega^2\, b\\
1&b&0&\omega^2 b&\omega b\\
1&\omega\, b&\omega^2\,b&0&b\\
1&\omega^2\,b&\omega b&b&0
\end{array}\right]\label{conf5}
\end{eqnarray}
which depends on two complex parameters, a discrete parameter $\omega$ that is solution of the equation $\omega^2+\omega+1=0$, and a free parameter  $b\in$\,\C$^*$.  
By using the same procedure as before one finds
\begin{eqnarray}
 O_{10}^{(5)}=
\left[\begin{array}{cccccccccc}
1&1&1&1&1&1&1&1&1&1\\
1&ab&a&a\omega&a\omega^2&-ab&-1&b&b/\omega& b/\omega^2\\
1&a&ac&a\omega^2&a\omega& -ac&c&-1&c/\omega^2&c/\omega \\
1&a\omega &a \omega^2&ad&a&-ad&d/\omega &d\omega^2 &-1&d\\
1&a\omega^2 &a\omega &a&ae&-ae&e/\omega^2 &e/\omega &e& -1\\
1&a\omega^2 &a\omega &a&-ae&ae&-e/\omega^2 &-e/\omega &-e& -1\\
1&-ab&a&a\omega&a\omega^2&ab&-1&-b&-b/\omega&- b/\omega^2\\
1&a&-ac&a\omega^2&a\omega& ac&-c&-1&-c/\omega^2&-c/\omega \\
1&a\omega &a \omega^2&-ad&a&ad&-d/\omega &-d/\omega^2 &-1&-d\\
1&-1&-1&-1&-1&-1&1&1&1&1
\end{array}\right]\label{O10}
\end{eqnarray}
 Thus if in   formula (\ref{O10}) one take $a,\,b,\,c,\,d,\,e \in\,$\T$^5$ one gets the Hadamard matrix
\begin{eqnarray}
D_{10}^{(5)}(\omega,a,b,c,d,e)=
 H_{10}^{(\omega)}\circ  EXP\left( i\cdot R_{10}^{(5)}(a,b,c,d,e)\right)\label{affine8}
\end{eqnarray}
where $R_{10}^{(5)} $ in (\ref{affine8})  has the form (\ref{R10})
with  $a,\,b\,,c,\,d,e\in\R $
\begin{eqnarray}
 R_{10}^{(5)}=
\left[\begin{array}{cccccccccc}
\bullet&\bullet&\bullet&\bullet&\bullet&\bullet&\bullet&\bullet&\bullet&\bullet\\ 
\bullet&a+b&a&a&a&a+b&\bullet&b&b&b\\
\bullet&a&a+c&a&a&a+c&c&\bullet&c&c\\
\bullet&a&a&a+d&a&a+d&d&d&\bullet&d\\
\bullet&a&a&a&a+e&a+e&e&e&e&\bullet\\
\bullet&a&a&a&a+e&a+e&e&e&e&\bullet\\
\bullet&a+b&a&a&a&a+b&\bullet&b&b&b\\
\bullet&a&a+c&a&a&a+c&c&\bullet&c&c\\
\bullet&a&a&a+d&a&a+d&d&d&\bullet&d\\
\bullet&\bullet&\bullet&\bullet&\bullet&\bullet&\bullet&\bullet&\bullet&\bullet
\end{array}\right]\label{R10}
\end{eqnarray}

and $ H_{10}^{(\omega)} $is given by
\begin{eqnarray}
 H_{10}^{(\omega)}=
\left[\begin{array}{rrrrrrrrrr}
1&1&1&1&1&1&1&1&1&1\\
1&1&1&\omega&\omega^2&-1&-1&1&1/\omega&/1\omega^2\\
1&1&1&\omega^2&\omega&-1&1&-1&1/\omega^2&1/\omega\\
1&\omega&\omega^2&1&1&-1&1/\omega&1/\omega^2&-1&1\\
1&\omega^2&\omega&1&1&-1&1/\omega^2&1/\omega&1&-1\\
1&\omega^2&\omega&1&-1&1&-1/\omega^2&-1/\omega&-1&-1\\
1&-1&1&\omega&\omega^2&1&-1&-1&-1/\omega&-1/\omega^2\\
1&1&-1&\omega^2&\omega&1&-1&-1&-1/\omega^2&-1/\omega\\
1&\omega&\omega^2&-1&1&1&-1/\omega&-1/\omega^2&-1&-1\\
1&-1&-1&-1&-1&-1&1&1&1&1
\end{array}\right]\nonumber
\end{eqnarray}

The above matrix is a generalized Butson matrix. Other such matrices can be obtained from (\ref{O10}) allowing the five parameters $(a,\,b,\,c,\,d,\,e)$ to take values from the set $(1,\,\omega,\,\omega^2)$.

Other known results in this dimension  are  those found from 10-dimensional Fourier matrix by Tadej  and $\dot{\rm Z}$yczkowski \cite{TZ}, which depends on four parameters, and that found by Sz\"oll\H{o}si \cite{FS} which depends on three parameters.

 By using the doubling formula (\ref{h16}) with $A$ and $B$ of the form $D_{10}^{(5)}$ and  $D\in$ \T$^{10}$  one gets a formula for 
 $D_{20}^{(19)}$ which depends on 19 arbitrary phases, and either $\omega=e^{2\pi{\bf{i}}/3}$, or $\omega=e^{4\pi{\bf{i}}/3}$.

For $n=6$   there are more different inverse complex conference matrices that are equivalent after re-parametrization. Taking into account   the results from \cite{D2} it follows that one matrix and its transpose are also equivalent. A consequence of this fact is   the conclusion that even we make  use of  the non-linear formula (\ref{conf}) one gets only \textit{one} matrix, instead of many as is asserted in \cite{D1}. Thus in the following we use only the matrix

\begin{eqnarray}
C_{6}=
\left[\begin{array}{cccccc}
0&1&1&1&1&1\\
1&0&b &b&-b &-b \\
1&b &0&-b&b/c&-b/c\\
1&b &-b&0&-b/c &b/c \\
1&-b&b c&-b\,c &0&b\\
1&-b &-b\,c &b\,c &b&0\\
\end{array}
\right]\label{conf6A}
\end{eqnarray}

It is easy to see that the above construction can be generalised such that the following proposition holds:
\newtheorem{pro}{Proposition}
\begin{pro}
Let us suppose that the  $n\times n$  complex inverse orthogonal conference matrix $C_n$ has the standard form, i.e. all the entries on the first row and the first column are equal to unity, excepting that on the main diagonal which is zero.  If   $C_n$ depends on $p$ complex parameters, then the complex inverse orthogonal  matrix $O_{2n}$ obtained by using formula (\ref{orth}) depends on  $n+p-1$ complex parameters. 
\end{pro}

The matrix (\ref{conf6A}) leads to the inverse orthogonal matrix (\ref{O12})  which depends on seven complex non-zero parameters, and to complex Hadamard matrices of the form (\ref{affine8}).

\begin{eqnarray}
 O_{12}^{(7)}(a,b,c,d,e,f,g)=~~~~~~~~~~~~~~~~~~~~~~~~~~~~~~~~~~~~~~~~~~~~~~~~~~~~~~~~~~~~~~~~~~~\nonumber\\
\left[\begin{array}{crrrrrrrrrrr}
1&1&1&1&1&1&1&1&1&1&1&1\\
1&ab&a&a&-a&-a&-ab&-1&b&b&-b&-b\\
1&a&ac&-a&a/g&-a/g&-ac&c&-1&-c&c/g&-c/g\\
1&a&-a&ad&-a/g&a/g&-ad&d&-d&-1&-d/g&d/g\\
1&-a&ag&-ag&ae&a&-ae&-e&eg&-eg&-1&e\\
1&-a&-ag&ag&a&af&-af&-f&-fg&fg &f&-1\\
1&-a&-ag&ag&a&-af&af&f&fg&-fg &-f&-1\\
1&-a&ag&-ag&-ae&a&ae&e&-eg&eg&-1&-e\\
1&a&-a&-ad&-a/g&a/g&ad&-d&d&-1&d/g&-d/g\\
1&a&-ac&-a&a/g&-a/g&ac&-c&-1&c&-c/g&c/g\\
1&-ab&a&a&-a&-a&ab&-1&-b&-b&b&b\\
1&-1&-1&-1&-1&-1&-1&1&1&1&1&1
\end{array}
\right]\label{O12}
\end{eqnarray}

The corresponding    Hadamard matrix  has the form

\begin{eqnarray}
D_{12}^{(7)}(a,b,c,d,e,f,g)= H_{12}\circ  EXP\left( i\cdot 
R_{12}^{(7)}(a,b,c,d,e,f,g)\right)\label{affine3}
\label{affine4}
\end{eqnarray}
where

\begin{eqnarray}
 H_{12}=
\left[\begin{array}{rrrrrrrrrrrr}
1&1&1&1&1&1&1&1&1&1&1&1\\
1&1&1&1&-1&-1&-1&-1&1&1&-1&-1\\
1&1&1&-1&1&-1&-1&1&-1&-1&1&-1\\
1&1&-1&1&-1&1&-1&1&-1&-1&-1&1\\
1&-1&1&-1&1&1&-1&-1&1&-1&-1&1\\
1&-1&-1&1&1&1&-1&-1&-1&1&1&-1\\
1&-1&-1&1&1&-1&1&1&1&-1&-1&-1\\
1&-1&1&-1&-1&1&1&1&-1&1&-1&-1\\
1&1&-1&-1&-1&1&1&-1&1&-1&1&-1\\
1&1&-1&-1&1&-1&1&-1&-1&1&-1&1\\
1&-1&1&1&-1&-1&1&-1&-1&-1&1&1\\
1&-1&-1&-1&-1&-1&-1&1&1&1&1&1
\end{array}
\right]\label{H12}
\end{eqnarray}

\begin{eqnarray}
R_{12}^{(7)}(a,b,c,d,e,f,g)=~~~~~~~~~~~~~~~~~~~~~~~~~~~~~~~~~~~~~~~~~~~~~~~~~~~~~~~~~~~~~~~~~~~~~~~~~~~~~~~~\nonumber\\
 \left[\begin{array}{cccccccccccc}
\bullet&\bullet&\bullet&\bullet&\bullet&\bullet&\bullet&\bullet&\bullet&\bullet
&\bullet&\bullet\\
\bullet&a+b&a&a&a&a&a+b&\bullet&b&b&b&b\\
\bullet&a&a+c&a&a-g&a-g&a+c&c&\bullet&c&c-g&c-g\\
\bullet&a&a&a+d&a-g&a-g&a+d&d&d&\bullet&d-g&d-g\\
\bullet& a&a+g&a+g&a+e&a&a+e&e&e+g&e+g&\bullet&e\\
\bullet&a&a+g&a+g&a&a+f&a+f&f&f+g&f+g&f&\bullet\\
\bullet&a&a+g&a+g&a&a+f&a+f&f&f+g&f+g&f&\bullet\\
\bullet& a&a+g&a+g&a+e&a&a+e&e&e+g&e+g&\bullet&e\\
\bullet&a&a&a+d&a-g&a-g&a+d&d&d&\bullet&d-g&d-g\\
\bullet&a&a+c&a&a-g&a-g&a+c&c&\bullet&c&c-g&c-g\\
\bullet&a+b&a&a&a&a&a+b&\bullet&b&b&b&b\\
\bullet&\bullet&\bullet&\bullet&\bullet&\bullet&\bullet&\bullet&\bullet&\bullet
&\bullet&\bullet
\end{array}
\right]\label{R12}
\end{eqnarray}
where we put the above relations in the most symmetric form.

If in relation (\ref{O12}) one take $a=i$ and $b=c=d=e=f=g=1$ one finds

\begin{eqnarray}
D_{12}=
\left[\begin{array}{rrrrrrrrrrrr}
1&1&1&1&1&1&1&1&1&1&1&1\\
1&{\bf i}&{\bf i}&{\bf i}&-{\bf i}&-{\bf i}&-{\bf i}&-1&1&1&-1&-1\\
1&{\bf i}&{\bf i}&-{\bf i}&{\bf i}&-{\bf i}&-{\bf i}&1&-1&-1&1&-1\\
1&{\bf i}&-{\bf i}&{\bf i}&-{\bf i}&{\bf i}&-{\bf i}& 1&-1&-1&-1&1\\
1&-{\bf i}&{\bf i}&-{\bf i}&{\bf i}&{\bf i}&-{\bf i} &-1&1&-1&-1&1\\
1&-{\bf i}&{-\bf i}&{\bf i}&{\bf i}&{\bf i}&-{\bf i}&-1&-1&1&1&-1\\
1&-{\bf i}&{-\bf i}&{\bf i}&{\bf i}&-{\bf i}&{\bf i} &1&1&-1&-1&-1\\
1&{-\bf i}&{\bf i}&{\bf -i}&{\bf -i}&{\bf i}&{\bf i}&1&-1&1&-1&-1\\
1&{\bf i}&{-\bf i}&-{\bf i}&{\bf -i}&{\bf i}&{\bf i}&-1&1&-1&1&-1\\
1&{\bf i}&-{\bf i}&{\bf -i}&{\bf i}&{\bf- i}&{\bf i}& -1&-1&1&-1&1\\
1&-{\bf i}&{\bf i}&{\bf i}&-{\bf i}&{-\bf i}&{\bf i} &-1&-1&-1&1&1\\
1&-1&-1&-1&-1&-1&-1&1&1&1&1&1
\end{array}
\right]\label{d12} \end{eqnarray}
which for this dimension $D_{12}$  is a novel matrix in literature.

\section{Conclusion}

In contradistinction to what we stated in the previous form of the paper, \cite{D1}, the number of independent parameters is smaller with two units in all the cases. Another important difference is the fact that by direct computation it was proved that a matrix and its transpose carry the same information such that in each case discussed in the paper one gets only a single matrix, relevant in this sense being the equivalence between $D_8^{(4)}$ and  $S_8^{(4)}$ found by  Sz\"oll\H{o}si, \cite{FS1}. This shows that  the number of nonequivalent Hadamard matrices could diminish in many cases and point out the difficulty of  the equivalence problem.

In this paper we provided a procedure to find parametrization of complex inverse orthogonal matrices by doubling the size of complex inverse orthogonal conference matrices. Thus a main problem is how to find them. If $n=2p$, $p=2,\,3,\dots$ one may use Paley construction \cite{RP} to construct (real) symmetric and skew-symmetric conference matrices. After that one has to check how many $\pm 1$ entries can be changed into complex non-zero numbers. When $n$ is odd the problem is  more difficult. If in formula (\ref{conf}) one takes $b=1$, $C_5$ matrix gets a symmetric conference matrix. Thus an idea is to look for such matrices and afterwards to proceed as in the even case.

Our approach can be used to obtain $n\times n$ complex weighing matrices, $W_{n,k}$, which are  similar to orthogonal conference matrices  (\ref{con}), being complex matrices that have on each row and column exactly $k$ zero entries and satisfy the relation
\begin{eqnarray}
W_{n,k}\,W_{n,k}^*=(n-k)I_n\label{weigh}
\end{eqnarray}
where $k$ is called weight. When $k=1$ one gets a conference matrix. 

It is easily seen that such a matrix can be brought to the form $(W_{n,k})_{ii}=0$, such that the doubling formula (\ref{orth}) can be used if the $W_{n,k}^{-1}$ entries are given by the following formula 
\begin{eqnarray}
(W_{n,k}^{-1})_{ij}=\left\{\begin{array}{cc}
1/(W_{n,k})_{ji}&\,{\rm for}\,\,(W_{n,k})_{ji}\ne 0\\
0&\,{\rm for}\,\,(W_{n,k})_{ji}= 0\end{array}\right.
\end{eqnarray}
The obtained matrix will be again a weighing matrix whose weight is now $2(k-1)$; for $k=1$ one get Hadamard matrices, for $k=2$ one get  weighing matrices with the same number of zero entries but with a double size, and for $k\ge 3$ one find  sparse matrices of the form $W_{2n,2(k-1)}$. These matrices are important for  applications in quantum information theory, cryptography, signal processing and orthogonal code design, see for example the paper \cite{ZL}. 

The weighing matrices for $k\ge 2$ can be obtained very easily, and here we give an example of  a complex  orthogonal $W_{4,2}$  
\begin{eqnarray}
W_{4,2}=\left[\begin{array}{cccc}
0&a&0&b\\
c&0&d&0\\
0&e&0&-\frac{b e}{a}\\
f&0&-\frac{df}{c}&0
\end{array}\right]\end{eqnarray}
which depends on six arbitrary non-zero parameters. Applications of weighing matrices will be given elsewhere.

\vskip3mm
{\bf Acknowledgments}
\vskip3mm
I would like to thank the referee for a few  remarks concerning the   equivalence problem for matrices obtained by restricting a set from all arbitrary  parameters to numerical value $1$, and by asking the question if there are known methods for finding complex conference matrices,  that lead to an improvement of the manuscript. I am much obliged to F  Sz\"oll\H{o}si who pointed out  that in my paper \cite{D1} not all the parameters are independent, and in this form I corrected the error. In fact in all cases their  number is n-2 of the old parameters.  I  thank also  M H Lee and  K ${\dot{\rm Z}}$yczkowsky for their interest on the topics discussed in the paper, and ANCS for partial support under project PN 09 37 0102

\end{document}